\documentclass[sigconf,nonacm]{acmart}
\usepackage{makecell}
\usepackage{siunitx}
\usepackage{listings}
\usepackage{subcaption}
\usepackage{cleveref}

\newcommand{\per}[1]{#1\,\%}

\makeatletter
\let\orig@lstnumber=\thelstnumber

\newcommand\lstsetnumber[1]{\gdef\thelstnumber{#1}}
\newcommand\lstresetnumber{\global\let\thelstnumber=\orig@lstnumber}
\makeatother

\begin{document}

\title{An Exploratory Study of Ad Hoc Parsers in Python}
\titlenote{Accepted as a registered report for MSR 2023 with Continuity Acceptance (CA).}

\author{Michael Schröder}
\affiliation{%
  \institution{TU Wien}
  \city{Vienna}
  \country{Austria}}
\email{michael.schroeder@tuwien.ac.at}
\orcid{0000-0003-1496-0531}

\author{Marc Goritschnig}
\affiliation{%
  \institution{TU Wien}
  \city{Vienna}
  \country{Austria}}
\email{marc.goritschnig@student.tuwien.ac.at}

\author{Jürgen Cito}
\affiliation{%
  \institution{TU Wien}
  \city{Vienna}
  \country{Austria}}
\email{juergen.cito@tuwien.ac.at}

\begin{abstract}
  \paragraph{Background:}
  Ad hoc parsers are pieces of code that use common string functions like \texttt{split}, \texttt{trim}, or \texttt{slice} to effectively perform parsing.
  Whether it is handling command-line arguments, reading configuration files, parsing custom file formats, or any number of other minor string processing tasks, ad hoc parsing is ubiquitous---yet poorly understood.

  \paragraph{Objective:}
  This study aims to reveal the common syntactic and semantic characteristics of ad hoc parsing code in real world Python projects.
  Our goal is to understand the nature of ad hoc parsers in order to inform future program analysis efforts in this area.

  \paragraph{Method:}
  We plan to conduct an exploratory study based on large-scale mining of open-source Python repositories from GitHub.  
  We will use program slicing to identify program fragments related to ad~hoc parsing and analyze these parsers and their surrounding contexts across 9~research questions using 25~initial syntactic and semantic metrics.
  Beyond descriptive statistics, we will attempt to identify common parsing patterns by cluster analysis.
\end{abstract}


\maketitle

\begin{figure}
  \vspace{1.5em}
  \input{fig-examples.tex}
  \vspace{-0.5em}
  \caption{Examples of ad~hoc parsers found on GitHub.}
  \label{fig:examples}  
  \vspace{-1em}
\end{figure}

\section{Introduction}

Ad~hoc parsers are everywhere, yet they go largely unnoticed.
We usually think of parsers as well-defined functions from strings to some other data type, based on more-or-less formally specified grammars; they often are significant standalone components of applications like compilers or web browsers.
Ad~hoc parsers, in contrast, are small, intermixed with business logic, and lack any formal specifications of their their input languages.

\Cref{fig:examples} shows some examples of ad~hoc parsers found in ordinary Python code.
An ad~hoc parser arises as soon as a string is processed in any way, whether through functions like \texttt{split} or \texttt{trim}, or even just by indexing into the string using common subscript notation like \texttt{s[i]}.
Although deceptively simple, all of these operations induce constraints on the string they are acting on, based on their specifications;
if this implicit contract is broken, the program will go wrong in some way.

Surprisingly, ad~hoc parsers are not very well studied.
Despite longstanding concerns about the security risks of ad~hoc input handling~\citep{bratus2014beyond}---and design patterns to avoid those risks~\citep{bratus2017usenix}---we know very little about the syntactic and semantic characteristics of actual ad~hoc parsing code in the wild.
We suspect ad~hoc parsers are scattered throughout codebases in a shotgun manner~\citep{momot2016seven}, but perhaps there are certain code markers around which ad~hoc parsing code tends to congregate?
We know ad~hoc parsers can be small, but what exactly are their typical sizes?
What functions do ad~hoc parsers typically call, what language features do they employ?
If they do handle errors, how do they go about it?
How complex is ad~hoc parsing code?
Is it amenable to static analysis?

The goal of this study is to shed light on how ad~hoc parsers operate and how they are utilized.
We want to inform future program analysis efforts in this area and are specifically motivated by concrete plans to infer grammars for ad~hoc parsers~\citep{schroeder2022grammars}, which require a solid empirical foundation.
In order to scope out suitable techniques for abstract interpretation and analysis, including precise abstract string domains, it is necessary to know the expected range and behavior of characteristics like loop bounds or exception-related control flow, among many others.

We chose an exploratory study design to survey a wide, partially unknown array of syntactic and semantic features of ad~hoc parsers and their surroundings.
In this first study, we focus on ad~hoc parsers in Python, a popular language for data science and machine learning tasks, which involve high amounts of text wrangling. 

\section{Research Questions}

\begin{itemize}
  \item[\bf RQ1] \textbf{How common are ad hoc parsers in Python?}\\
  First, we want to know how prevalent ad~hoc parsers are in the wild.
  We can determine this by looking at the number of projects that contain at least one ad~hoc parser, and the ratio of ad~hoc parsing code to all other code in a project.
 
  \smallskip
  \item[\bf RQ2] \textbf{Where are ad hoc parsers located?}\\
  One might think that the parsing component of a function is typically at the beginning, validating and transforming inputs before they are passed on to the rest of the program.
  But we know that \emph{shotgun parsing}---the intermixing of parsing and business logic---is a real phenomenon~\citep{momot2016seven,underwood2016search}.
  We want to know how often this actually occurs on the function level.
  We also want to locate ad~hoc parsers on the system level:
  Do they only appear at the edges of a system, near I/O operations, or perhaps also deep within projects, where strings are used as a quick way to pass around semi-structured data?  
 
  \smallskip
  \item[\bf RQ3] \textbf{How large are ad hoc parsers?}\\
  By definition, ad~hoc parsers are small snippets of code, but we do not know what their actual average size is, in terms of lines of code or number of expressions.
  We do not know whether ad~hoc parsers regularly use temporary variables to store intermediate results or perhaps not use any variables at all, preferring method chaining.
  Ad~hoc parsers might be syntactically compact but also pack complex functionality in a small space.
  
  \smallskip
  \item[\bf RQ4] \textbf{What are the input sources of ad hoc parsers?}\\
  The immediate source of an ad~hoc parser's input string could be an argument of the enclosing function, a global or instance variable, or the return value of some function call.
  In many cases, we should be able to determine the ultimate origin of the input, e.g., a command-line argument (stored in \texttt{sys.argv}) or a line read from a file (via \texttt{readline}).
  
  \smallskip
  \item[\bf RQ5] \textbf{What functions do ad hoc parsers use and how?}\\
  We want to know exactly which common functions and operations make up a typical ad~hoc parser, and how they are used.
  One would certainly expect string functions like \texttt{split} or \texttt{strip} to feature prominently, but what about sequence operations like \texttt{map} or \texttt{index}, or syntactic sugar like \texttt{s[i:j]} for slicing?
  What are common arguments used with these functions?
  Do ad~hoc parsers in Python use tuples and multiple return values?
  Do they use non-standard user-defined functions, which could impact static analysis by increasing the call graph that has to be investigated, potentially even introducing non-local effects?
  
  \smallskip
  \item[\bf RQ6] \textbf{How do ad hoc parsers use regular expressions?}\\
  A characteristic of ad~hoc parsers is that they use common functions to parse strings, rather than more formal methods of parsing.
  Regular expressions, while ostensibly a proper formal parsing method, are nonetheless regularly used in an ad~hoc fashion.
  They are often combined with other parsing constructs and may only play a small part in a larger piece of parsing code.
  We want to know how often ad~hoc parsers use regular expressions internally and to what end.
  Previous investigations have focused on regular expressions in isolation~\citep{chapman2016regex,davis2018redos,davis2020regex}, but have not ventured into a more holistic inquiry on the combination of regular expressions and ad~hoc parsing.
  For example, are regular expressions used to do a first pass over the input string, using features such as named groups to break down the input's superstructure, before parsing continues on the smaller pieces? 
  Or are they used at the terminal point of the input language, i.e., do ad~hoc parsers first use functions like \texttt{split} and then apply regular expressions to the results?
  We want to know what kinds of regular expressions are used by ad~hoc parsers and whether the use of regular expressions within ad~hoc parsers produces non-regular languages, or whether the parser could have been written entirely as a regular expression (disregarding any readability concerns).
  This last question we will only be able to answer approximately, as we do not (yet) have a precise method of determining the input language of an ad~hoc parser.
  Certain heuristics, such as branching structure and the nature of any enclosing loop bounds, might give us some hints, however.
  
  \smallskip
  \item[\bf RQ7] \textbf{What is the nature of loops in ad hoc parsers?}\\
  Every parser will in some way loop over its input string to access the string's characters.
  This can be done in a high-level functional manner, using functions like \texttt{map} or \texttt{split}, or by directly iterating over characters, using \texttt{for} or \texttt{while} loops.
  Loops can also be used to iterate over substrings of the input string, e.g., the results of a use of \texttt{split}.
  Loops can be nested, and it is even possible that a parser involves a recursive call to the enclosing function.
  We want to assess how ad~hoc parsers use these various looping constructs and classify them accordingly.
  Of particular interest is the type of loop bound, as this will have a big impact on static analysis.
  Functions like \texttt{split} are always implicitly bounded by the length of their input, whereas other looping constructs allow for more complex bounds.
  
  \smallskip
  \item[\bf RQ8] \textbf{How do ad hoc parsers handle errors?}\\
  Every parser rejects those strings that are not part of the language it is parsing.
  In other words, a parser fails if it is fed an unknown string.
  How do ad~hoc parsers deal with this?
  Do they crash?
  Perhaps an exception is (implicitly) raised but caught by the enclosing function.
  Or perhaps the ad~hoc parser handles failure explicitly, returning an error value or a default value.
  How ad~hoc parsers handle exceptions is of utmost importance, as this determines whether or not they might pose a fault risk.
  
  \smallskip
  \item[\bf RQ9] \textbf{What are typical ad hoc parsing patterns?}\\
  Beyond compiling descriptive statistics about ad~hoc parsers, we want to identify particular patterns of parsing, perhaps even a taxonomy of ad~hoc parser types.
  Are there certain combinations of syntactic and semantic features that commonly co-occur?
  Can we identify certain application domains (based on identifier names and string origins) in which particular types of ad~hoc parsers occur more often?
  A set of ad~hoc parsing patterns would help researchers in talking about phenomena related to ad~hoc parsing
\end{itemize}

\begin{table*}
  \caption{Initial list of metrics extracted for each ad~hoc parser.}
  \label{tab:metrics}
  \vspace{-0.5em}
  \begin{tabular}{c@{\;}c@{\;}c@{\;}c@{\;}c@{\;}c@{\;}c@{\;}clp{0.66\linewidth}}
\toprule
\multicolumn{8}{c}{RQs} & Metric & Description \\
\midrule
1 & 2 &   &   &   &   &   &   & Project Name        & name of the project containing the ad hoc parser \\
1 &   &   &   &   &   &   &   & Project LOC         & total lines of code in the containing project \\
  & 2 &   &   &   &   &   &   & Module Name         & name of the enclosing module/file \\
  & 2 &   &   &   &   &   &   & EF Name             & name of the enclosing function \\
  & 2 &   &   &   &   &   &   & EF LOC              & total lines of code in the enclosing function \\
  & 2 &   &   &   &   &   &   & Position            & position of the ad hoc parser within the enclosing function \\
1 & 2 & 3 &   &   &   &   &   & LOC                 & lines of code in the ad hoc parser \\
  &   & 3 &   &   & 6 &   &   & CYCLO               & cyclomatic complexity of the ad hoc parser \\
  & 2 &   & 4 &   &   &   &   & Input Source        & source of the input string: EF argument, global variable, function call, etc. \\
  & 2 &   & 4 &   &   &   &   & Input Origin        & origin of the input string: command-line, file, environment variable, etc. \\
  &   & 3 &   &   &   &   &   & Expression Count    & number of expressions in the ad hoc parser \\
  &   & 3 &   &   &   &   &   & Variable Count      & number of variables in the ad hoc parser \\
  &   & 3 &   &   &   &   &   & Function Count      & number of function calls in the ad hoc parser \\
  &   &   &   & 5 & 6 & 7 & 8 & Function Names      & names of all functions called in the ad hoc parser \\
  &   &   &   & 5 &   &   &   & Function Origins    & origin of each called function: user-defined or from a library \\
  &   &   &   & 5 & 6 &   &   & Function Positions  & position of all function calls within the ad hoc parser \\
  &   &   &   & 5 &   &   &   & Function Arguments  & arguments with which each function is called, besides the input string \\
  &   &   &   & 5 &   &   & 8 & Syntactic Sugar     & special syntax used in the ad hoc parser: subscript notation, tuples, list comprehensions, etc. \\
  &   &   &   &   & 6 &   &   & Regular Expressions & arguments to known regex functions or regex literals used in the ad hoc parser \\
  &   &   &   &   & 6 & 7 &   & Loop Bounds         & constant, linear on input string, complex, or unbounded \\
  &   &   &   &   &   & 7 &   & Loop Types          & \texttt{for}, \texttt{while}, functional (\texttt{map}, \texttt{split}, etc.), or recursive \\
  &   &   &   &   & 6 & 7 &   & Loop Nesting Depth  & how deeply nested loops in the ad hoc parser are \\
  &   &   &   &   &   &   & 8 & Caught Exceptions   & all exceptions caught by the ad hoc parser or the enclosing function \\
  &   &   &   &   &   &   & 8 & Uncaught Exceptions & all uncaught exceptions (excluding explicitly raised ones) \\
  &   &   &   &   &   &   & 8 & Raised Exceptions   & all exceptions explicitly raised by the parser (using \texttt{raise}) \\
\bottomrule
\end{tabular}

\end{table*}

\section{Execution Plan}

\subsection{Dataset \& Infrastructure}

To collect and analyze a large-scale dataset of Python projects, we plan on using Boa~\citep{dyer2013boa}, a source code mining language and infrastructure.
Boa allows running static program analysis at scale, using a declarative domain-specific language with built-in support for complex analysis tasks such as control-flow graph (CFG) generation and traversal~\citep{dyer2013visitors}.
It has been previously used to extensively analyze syntactical features of Python programs~\citep{dyer2022python}, which gives us confidence in the feasibility of our envisioned analyses.

As of this writing, the latest Boa Python dataset (February 2022) includes \num{104424} GitHub projects that indicated Python as their primary language.
The repositories in the dataset were selected by sorting several million Python projects on GitHub by decreasing star count and decreasing date and thus reflect recent high-profile open-source Python projects (as of summer/fall 2021).\footnote{Robert Dyer, lead researcher on Boa, email to authors, March 13, 2023.}
The average star count in the dataset is 243 (min 24, median 59, max \num{138438}) and most projects (\per{55}) had commits within the last two years.

An advantage of using the Boa framework is that our analysis will be easily reproducible and can be applied to other datasets in the future.
As Boa is inherently a language-agnostic toolset, it should also be relatively easy to adapt our analysis to other programming languages, especially in comparison to custom one-off analysis scripts.

\subsection{Program Slicing}

To extract ad~hoc parsers from the dataset, we will use a form of program slicing~\citep{weiser1984program}, leveraging the built-in static analysis capabilities of the Boa framework.
Here is an outline of our approach:
\begin{enumerate}
  \item Extract all methods from all Python files in each project (including the top-level environment, which is treated like a regular method called \verb|__main__|).
  
  \item For each method, identify all string variables (including arguments).
  As Python is (usually) untyped, we have to perform crude but effective type inference by consulting an extensive list of methods whose arguments or return values are known to be (or not to be) strings, e.g., \texttt{split} or \texttt{startswith}.
  If type hints are available, we take those into consideration as well.
  While we might not be able to find strictly \emph{all} string variables of a method this way, we should be able to find most \emph{relevant} string variables, i.e., those involved in ad~hoc parsing.
  It seems highly unlikely that an ad~hoc parser would not use at least one unambiguously string-specific operation.
  
  \item For each string variable, construct a forward slice of the program, starting at the first occurrence of the variable (if it is not already part of a previous slice).
  We use an intra-procedural program-dependence graph (PDG)~\citep{ferrante1987pdg} to build the slice, continuing as long as the data dependents are themselves strings or collections of strings.
  This ensures that we capture the core of the parser, including intermediate results and transformations, but that we don't end up with a slice the size of the whole method.
  Our slices never extend beyond function boundaries.

  \item If a program slice does not include any methods that impose constraints on the input string (e.g., if the string is just repeatedly appended to), it is not a parser and therefore discarded.
\end{enumerate}

The program slices collected in this way capture the core of each ad~hoc parser, beginning with the appearance of the input string and ending at the point where no more transformations of that string or its substrings occur.
The parsed data types might be constrained further downstream, e.g., a parsed integer might be required to fall within a certain range, thus introducing further constraints on the input, but that is outside the scope of our present study.
While the delineation of ad~hoc parsing and business logic is fluid---a defining characteristic of ad~hoc parsing---we want to focus purely on the initial string parsing aspects.

\subsection{Analysis}

We will use the abstract syntax trees (ASTs) of the ad~hoc parser cores extracted using program slicing as the basis of our analysis.
For questions that require we look at the surrounding context, or at variables referenced by the core but not part of it (e.g., loop bounds), we can traverse outside the core AST on-demand as necessary.

While for most of our research questions we envision performing large-scale quantitative analysis on the ASTs, we want to complement our investigation with qualitative methods where we anticipate limitations due to soundness and completeness of our program analyses.
Specifically, this means we will also sample ad~hoc parsers in source code form for manual inspection.

To answer \textbf{RQs 1--8}, we will extract a number of metrics from the ad~hoc parser ASTs and use them to generate various descriptive statistics.
\Cref{tab:metrics} shows an initial but not exhaustive list of these metrics.
As this is an exploratory study, we anticipate that additional opportunities for insight will arise as we survey the data and thus we are prepared to extend our efforts beyond the pre-defined metrics.

To answer \textbf{RQ9}, we will attempt to cluster the collected ad~hoc parsers based on the extracted metrics.
We will experiment with using $k$-means as a baseline for clustering, followed by more advanced learning methods leveraging higher-dimensional embeddings~\cite{xie2016unsupervised}.
We plan to experiment with different concrete code embedding methods, such as code2vec~\cite{code2vec}, which represents code snippets as single fixed-length code vectors; CoCLuBERT~\cite{coclubert}, a fine-tuned version of CuBERT~\cite{cubert} designed for code clustering; and inst2vec~\cite{inst2vec}, which defines an embedding space based on an intermediate representation of code.
We will then manually sample parsers from the identified clusters, both to validate the clustering and to gain further insight into the nature of the identified cluster.

\section{Threats to Validity}

\paragraph{Internal Validity.}

We use an established large-scale dataset of open-source Python projects collected from GitHub as the basis of our analysis.
It is possible that this dataset is not representative of Python code (and thus ad~hoc parsers in Python) at large.
To mitigate this risk, our entire analysis pipeline will be written in a reusable manner, running on the Boa infrastructure, which will allow future researchers to easily replicate our study on different and larger datasets.

\paragraph{External Validity.}
In this study, we only consider ad~hoc parsers in Python.
The characteristics of these parsers might be (partially) Python-specific, and thus might not generalize to ad~hoc parsers in other programming languages.
However, even if that were the case, the results of this study are still valuable for program analysis efforts within the Python ecosystem.

\paragraph{Construct Validity.}
Our program slicing method might be unsound or incomplete, capturing irrelevant program fragments or missing out on (parts of) some legitimate ad~hoc parsers.
To mitigate this risk, we combine our quantitative analysis methods with qualitative investigations, which allows us to validate the program slicing results by directly inspecting the original sources.

\section{Preliminary Study}

In a preliminary study, we collected and analyzed \num{12632} Python \verb|from_string| methods from open-source projects on GitHub.
We chose \verb|from_string| methods as a proxy for ad~hoc parsers, as these are small single-purpose functions that transform strings, usually originating in files, to internal data types.

We found that more than half of these ad~hoc parsers are less than 11 lines of code in size, with only \per{20} exceeding 20 lines, and that \per{95} have a cyclomatic complexity of at most 10.
The average number of functions called within a parser is 6, the median 3, and the most common operation is \texttt{split}, occurring in \per{41} of all parsers, followed by \texttt{len} and the \texttt{int} constructor, each occurring in about \per{29} of parsers.
Only \per{12} contain loops bounded by the length of the input string, \per{2} loops with other types of bounds, and \per{2} completely unbounded loops.
More than half of all parsers (\per{57}) have the potential to raise exceptions based on the operations they use (e.g., the \texttt{index} function on strings, which raises an exception when the given substring is not found) and almost half of those (\per{45}) due to the implicit possibility of out-of-bounds errors, i.e., unchecked array access or optimistic tuple assignment, which occurs when a function call has the potential to return a different number of variables than a tuple assignment syntactically expects (\per{28} of split operations are immediately followed by a tuple assignment).
Of all exception-raising parsers, \per{26} do so explicitly, using the \texttt{raise} keyword, and \per{11} of all investigated parsers explicitly catch and handle exceptions within the \verb|from_string| method.

These preliminary results give us an initial impression of ad~hoc parser characteristics but are limited by the fact that they are exclusively derived from \verb|from_string| methods.
While these are an interesting programming pattern in itself, we suspect that the kind of ad~hoc parsing happening in these methods is not necessarily generalizable.
By virtue of being so clearly delimited into their own functions, the parsers constituting \verb|from_string| methods do not exhibit the intermixing of parsing with other code, which we think is a typical characteristic of ad~hoc parsers.
With the proposed study, we want to extend the scope of our inquiry to capture the phenomenon of ad~hoc parsing at large.

\bibliographystyle{ACM-Reference-Format}
\bibliography{references}

\end{document}